\documentstyle[amssymb]{elsart}
\begin{document}
\input epsf
\epsfclipon
\begin{frontmatter}
\title{ A simple model of bank bankruptcies }
\author[pw]{Agata Aleksiejuk\thanksref{emaa}} and
\author[pw,ge]{Janusz A. Ho\l yst\thanksref{emjh}}
\address[pw]{Faculty of Physics, Warsaw University
of Technology, \\ \hspace*{0.15em} Koszykowa 75, PL-00-662 Warsaw, Poland}
\address[ge]{Institute for Economics and Traffic, Dresden University
of Technology, Andreas-Schubert-Stra\ss e 23, D-01062 Dresden, Germany }
\thanks[emaa]{{\it E-mail}: agatka@if.pw.edu.pl}
\thanks[emjh]{Corresponding author. Fax: +48-22-6282171; {\it E-mail}: jholyst@if.pw.edu.pl}

\date{\today}

\begin{abstract}
Interbank deposits (loans and credits) are quite common in banking system all
over the world. Such interbank co-operation is profitable for banks but it can
also lead to collective financial failures. In this paper we introduce a new
model of directed percolation as a simple representation for contagion process
and mass bankruptcies in  banking systems. Directed connections that are
randomly distributed between junctions of bank lattice simulate flows of money
in our model.  Critical values of a mean density of interbank connections as
well as static and dynamic scaling laws for the statistic of avalange
bankruptcies are found. Results of computer simulations  for the universal
profile of bankruptcies spreading are in a qualitative agreement with the third
wave of bank suspensions during The Great Depression in the USA.
\end{abstract}


\end{frontmatter}



\section{Introduction}
Making a short review of latest publications on the percolation phenomenon one
could come to the conclusion that the percolation theory \cite{1} is a universal
paradigm  for physics, sociology and economy. In fact, percolating systems
composed of large number of interacting units can be simply adopted for
simulations of complex behaviours and environments. Number of such adoptions
have been done so far including microscopic simulations of the stock market
\cite{2}, social percolation models \cite{3} and marketing percolation
describing diffusion of innovations \cite{4}. Here we propose a simple model
basing on the intuitive similarity between percolation and banking networks.

At present economists are in agreement that the robustness of a country
financial system is related to the strength of a domestic economy. Bank
bankruptcies usually follow dramatic changes in the banking capital, assets as
well as liabilities and can be socially costly. In general two factors may cause
a bank failure: bad credits and rapid withdrawing of deposits. Economic
researches confirm that solvent and insolvent banks alike can experience
withdrawals for reasons unrelated to the bank failure risk in circumstances of a
banking panic \cite{5}. The same investigations emphasize the importance of
withdrawals rates. In fact, sudden withdrawals can have dramatic effects on the
bank stability and may force a bank to bankruptcy in a short time if it does not
receive assistance from other banks. On the other hand a bankruptcy of a single
bank can start  an avalanche of other bank failures due to the {\it domino
effect}.

\section{The model}
In our model banks are represented by vertices in a lattice that
for simplicity has a square or cubic symmetry. Directed
connections that are randomly distributed between banks simulate
flows of money. Banking capital consists of assets and liabilities
as in reality. Arrows entering into vertices represent liabilities
(deposits of other banks). Branches with opposite direction
reflect assets (investments and given credits). It follows that an
average number of arrows entering into a vertex is equal to an
average number of exiting arrows. We assume that even one
withdrawal or bad credit can force the bank  bankruptcy and one
failing bank can cause bankruptcies of other banks. Only {\it
interbank} credit connections  are considered, i.e.  bank deposits
and investments are neglected and no insurance system is assumed
in our model.

\begin{figure}
\epsfxsize=14cm \epsfbox{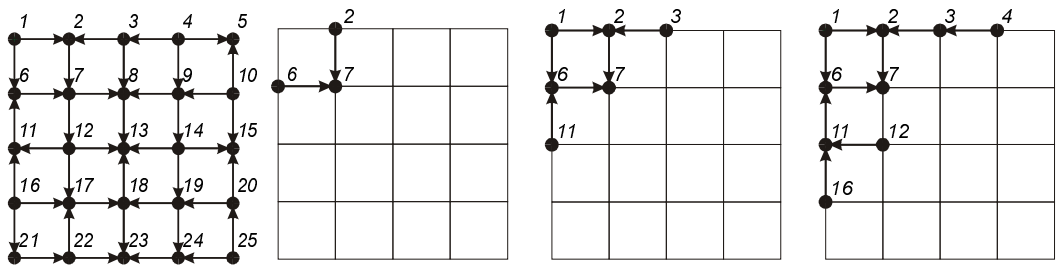} \caption{Bankruptcy spreading in banking
network based on square lattice.} \label{fig1}
\end{figure}

Dynamical rules governing time evolution of the model are as follows. Initially
each bank is solvent. The first bankrupt is selected at random and we do not
specify the reason for this bankruptcy that can be a bad credit or sudden
deposit withdrawal. During the next time step neighbouring banks loss their
solvency if they gave a loan to the bankrupt. This process is repeated until no
bank survives that gave a bad interbank credit. Above mentioned rules become
comprehensible after tracing Fig.~\ref{fig1}. The figure presents a  system with
$N=25$ banks. All possible flows of money (connections between vertices) are
realized in this pattern. Let us choose the 7th vertex as the first bankrupt.
According to rules assumed earlier the collapse of this bank forces suspension
of two other banks with numbers $\{2,6\}$. During the next step three other
banks are swept $\{1,3,11\}$. At the end, the avalanche originating from the
bank with the number $7$ includes nine banks $\{1,2,3,4,6,7,11,12,16\}$.

Despite the seeming similarity of our model to the well known {\it
directed percolation} \cite{1} it is based on a  new approach to
this phenomenon. In the traditional directed percolation
directions in space are not equal  i.e. the system is anisotropic
and  one direction, which is called the growth direction, is
special. In our model all directions are equal. There is also
another feature distinguishing the presented model from the
standard percolation. In both cases of the traditional site and
bond percolation each occupied site/bond belongs to {\it only one}
cluster. In our model this condition is not valid and the same
bank can be included to {\it various} avalanches depending on the
first bankrupt.

\section{Computer simulations}
We investigated statistics of bankrupt avalanches in systems
characterized by different mean concentrations $p$ of existing
interbank deposits. It follows that the system parameter  is the
same as in the  percolation theory. In analogy to the traditional
percolation one can expect  a critical value  $p_{c}$ when an
avalanche composed of bankrupts can spread {\it all over} the
banking network. This phenomenon is related to the percolation
phase transition. We performed numerical calculations in order to
estimate $p_{c}$ and basing on the finite size scaling law
$p(L)-p_{c}\sim L^{-\frac{1}{\nu}}$ (where $L$ is the linear size
of the system) we found that critical values $p_c$ in our model
are approximately two times larger than in the usual bond
percolation, i.e.   $p_c^{2D} \approx 1.00\pm0.01$ and $p_c^{3D}
\approx 0.51\pm0.02$   for the square and the cubic lattice
respectively.

\begin{figure}v
\epsfxsize=8cm \epsfbox{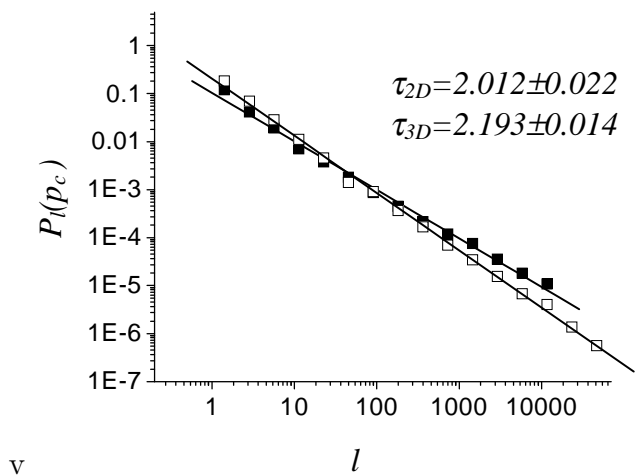} \caption{Avalanche length distribution at
$p_{c}$ in square lattice (solid squares) and cubic lattice (open
squares).}\label{fig2}
\end{figure}

\begin{figure}
\epsfxsize=8cm \epsffile{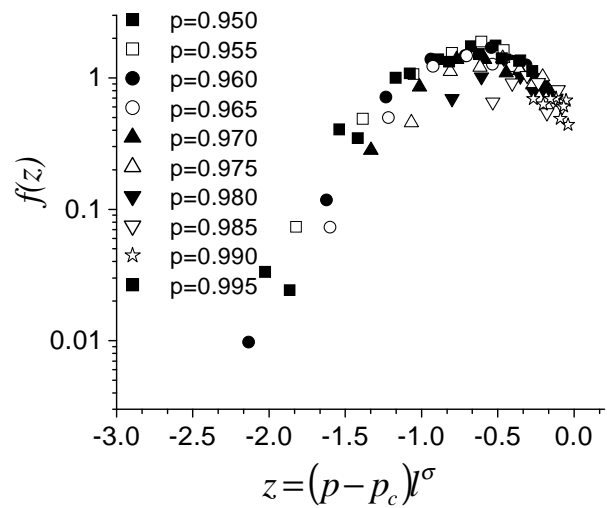} \caption{Scaling behaviour for the
renormalized avalanche statistics described by  $f(z)=P_l(p)/P_l(p_c)$.}
\label{fig3}
\end{figure}

We observed that distributions of avalanche lengths have the same
properties as statistics of cluster numbers in the usual
percolation system. At the percolation threshold the probability
that a random bank causes $l$-avalanche (failures of $l$ other
banks) fulfills the power law $P_{l}(p_{c})\sim l^{1-\tau}$ where
$\tau$ is the Fisher exponent. In both two and three dimensional
systems numerically calculated Fisher exponents are consistent
with their equivalents taken from the literature
(Fig.~\ref{fig2}). For $p$ near $p_{c}$ and for
$l\rightarrow\infty$ we found a good agreement with  the scaling
law describing avalanche distribution
$P_{l}(p)=l^{1-\tau}f[(p-p_{c})l^{\sigma}]$, where $f$ is a
universal scaling function. Fig.~\ref{fig3} illustrates this law
for a square lattice when the scaling exponent
$\sigma_{2D}=\frac{36}{91}$ has been used.

\begin{figure}
\epsfxsize=15cm \epsfbox{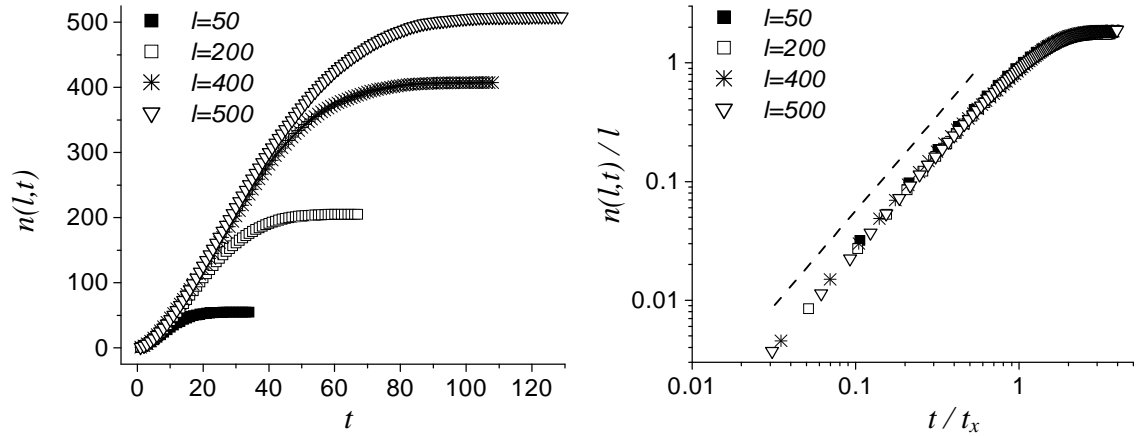} \caption{Time evolution of avalanche growth
in the square lattice with $L=512$ for different  avalanche  lengths $(l)$. The
number of banks that became bankrupts until the time $t$ is presented at the
vertical axis. Both right and left plots present the same data. Data on the
right plot correspond to  data from the left plot rescaled according to the
Family-Vicsek scaling relation.} \label{fig4}
\end{figure}

Dynamical properties of our model are described by the number of banks $n(l,t)$
swept during the bankruptcy avalanche up to the moment $t$ where $l$ is the
total avalanche length ($\lim_{t\rightarrow\infty} n(l,t)=l$). A typical plot of
$n(l,t)$  (Fig.~\ref{fig4}) has two regions separated by a {\it crossover time}
$t_{x}$ \cite{7}. Initially, when $t<<t_{x}$ the number of bankrupts increases
as $n(l,t)\sim t^{\beta}l^{\gamma}$. In the dynamic scaling theory of surface
growth the analogous exponent $\beta$ is called the {\it growth exponent}.
Fortunately for bank shareholders, the power-law increase is followed by the
saturation regime for $t>>t_{x}$. The saturation time $t_{x}$ depends on the
avalanche length as $t_{x}\sim l^{z}$. By analogy to the standard terminology
\cite{7} we call $z$ the {\it dynamic exponent}. We found that the avalanche
growth in our model fulfills the Family-Vicsek scaling relation $n(l,t)\sim l
g(t/l^{z})$ where $g$ is a universal scaling function (Fig.~\ref{fig4}) . The
scaling exponents $\beta$, $\gamma$, $z$ are connected by the equation
$\gamma+z\beta=1$. According to our numerical studies for the square lattice the
exponents account to $\beta=1.60\pm0.01$, $\gamma=0.08\pm0.02$, $z=0.56\pm0.01$
and do not depend on the system parameter $p$.

\begin{figure}
\epsfxsize=15cm \epsfbox{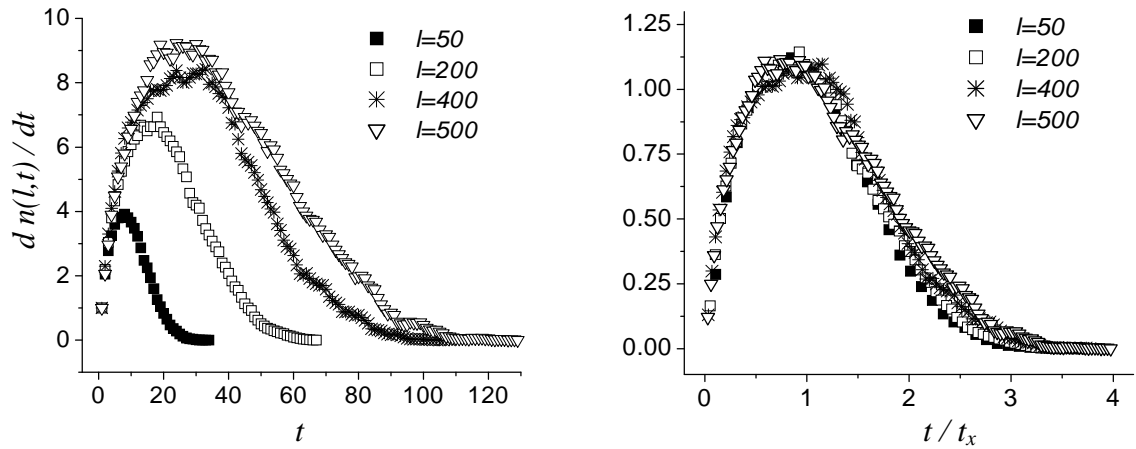} \caption{Speed of avalanche spreading for the
same data as in Fig.~\ref{fig4}. The right plot presents rescaled data.}
\label{fig5}
\end{figure}

Fig.~\ref{fig5} shows time distributions of bankruptcies belonging to avalanches
presented at Fig.~\ref{fig4}, i.e. the curves in Fig.~\ref{fig5} are  the first
derivatives of those in Fig.~\ref{fig4}. Observing the speed of avalanche
spreading we found a clear maximum which corresponds to the highest probability
of bankruptcy. Fig.~\ref{fig4} and Fig.~\ref{fig5} clearly show that there is a
unique mechanism governing avalanche growth in our model. The mechanism is
independent on the system parameter as well as on the avalanche length. Our
preliminary studies on cubic lattices prove that the same mechanism governs the
avalanche growth in three dimensional systems.

\begin{figure}
\epsfxsize=12cm \epsfbox{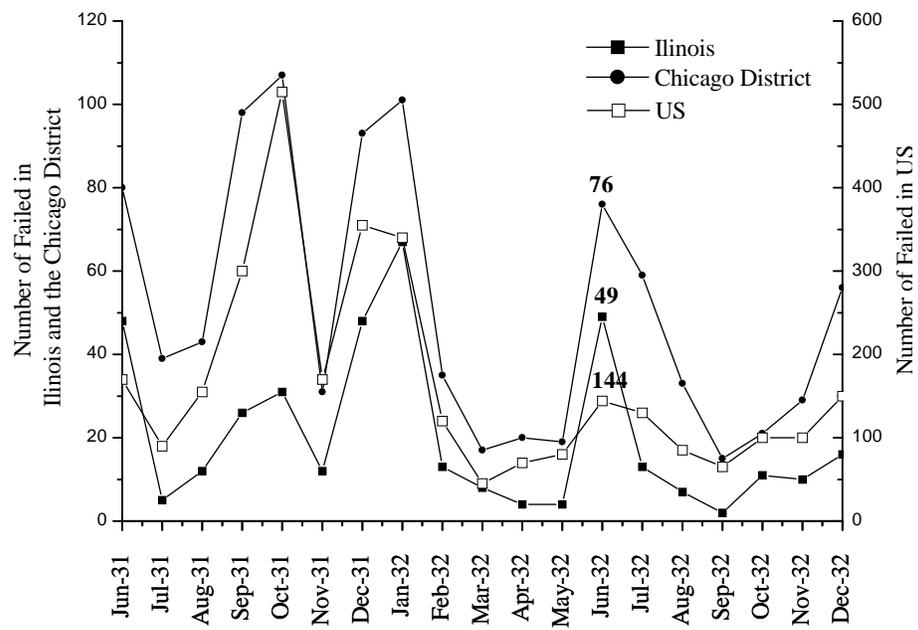} \caption{Total bank suspensions in Illinois,
the Chicago Federal Reserve District, and the US, monthly, June 1931-December
1932. After \cite{5}, courtesy of Ch.W. Calomiris and J.R. Mason.} \label{fig6}
\end{figure}

According to our knowledge this work is the first one connecting problems of
bank failures with the statistical physics. Although exact data concerning
spatial and time evolution of mass bankruptcies are hard to receive, it is known
that such bankruptcies were quite frequent in the nineteenth and  twentieth
century \cite{6}. The banking crisis that accompanied The Great Depression was
probably the most dramatic. Economists distinguish three waves of bank failures
during this period and the third wave (starting in May  1932) can be seen as
qualitatively consistent with our directed percolation model. In fact, during
the period May-Sep 1932 distributions of total bank suspensions in Illinois, the
Chicago Federal Reserve District and the USA  have shapes   (Fig. ~\ref{fig6})
similar to the time profile observed in our model (Fig.~\ref{fig5}). Contrary to
the situation during the earlier massive bank collapses in USA there was no
significant interventions from government institutions in order to stop the
contagion of banking system in this time \cite{5}.

At present government institutions guard security of banking system therefore
the black scenario known from The Great Depression seems incredible but it can
repeat. It is necessarily to emphases that the proposed model would be   more
realistic if it were widened to the whole financial system composed not only of
banks but also other financial institutions like trust or pension funds,
insurance companies and firms. Although each institution enumerated above
possesses  a different capital structure but all of them suffer from risks
related to bad investments/credits and are connected one to another.

\section{Conclusions}
 The model presented here has been thought to reflect the
cooperative behaviour of banking systems. We have shown that
avalanches of bankruptcies can be related to clusters in the
random directed percolation problem. It follows that  a large
number of interbank credits can lead to the percolation phase
transition when bankruptcies can spread all over the banking
network. Static and dynamic properties of this model are in a good
agreement with the percolation theory. The observed in numerical
simulations  shape  of avalanche spreading is in a qualitative
agreement with data from The Great Depression.

  Acknowledgements. One of us (JAH) is thankful to Prof. Dirk Helbing
   for his hospitality during the stay in Dresden. The work  has
   been in part supported by the ALTANA AG due to the Herbert Quandt-Programm.

\newpage


\end{document}